# Does the Mott problem extend to Geiger counters?

Jonathan F. Schonfeld

Center for Astrophysics | Harvard and Smithsonian

60 Garden St., Cambridge MA 02138 USA

jschonfeld@cfa.harvard.edu

ORCID ID# 0000-0002-8909-2401

**Abstract:** The Mott problem is a simpler version of the quantum measurement problem that asks: Is there a microscopic physical mechanism – based (explicitly or implicitly) only on Schroedinger's equation – that explains why a single alpha particle emitted in a single spherically symmetric s-wave nuclear decay produces a manifestly non-spherically-symmetric single track in a cloud chamber? I attempt here to generalize earlier work that formulated such a mechanism. The key ingredient there was identification of sites at which the cross section for ionization by a passing charged particle is near singular at ionization threshold. This near singularity arose from a Penning-like process involving molecular polarization in sub-critical vapor clusters. Here, I argue that the same Mott-problem question should be asked about Geiger counters. I then define a simple experiment to determine if ionization physics similar to the cloud chamber case takes place in the mica window of a Geiger counter and explains the collimation of wavefunctions that are spherically symmetric outside the counter into linear ion tracks inside. The experiment measures the count rate from a radioactive point source as a function of source-window separation. I have performed a proof-of-concept of this experiment; results are reported here and support the near-singular-ionization picture. These results are significant in their own right, but also because they may shed light on physical mechanisms underlying instances of the full quantum measurement problem. I illustrate this for the Stern-Gerlach experiment and a particular realization of superconducting qubits. I conclude by detailing further work required to flesh out these results more rigorously.

## 1. Introduction

The most conventional formulation of quantum mechanics [1] holds that a physical system is characterized by a wavefunction (Hilbert-space vector) which, *between measurements*, evolves smoothly and unitarily in time according to Schroedinger's equation, but which evolves discontinuously and non-unitarily at the moment of measurement. The quantity being measured is characterized by a Hermitian operator, and at the moment of measurement, the wavefunction is projected (collapses) onto a random eigenvector of this operator, with probability given by the absolute-value-squared of the inner product between the eigenvector in question and the state just prior to measurement, i.e. the Born rule.

This seems to say, counterintuitively, that Nature is thoroughly oblivious to even the finest details of experiments that, today, can be designed down to the atomic level. And yet, this picture, as far as it's been tested, has genuine empirical support, from interference experiments to qubit readout [2-7]. There have been many creative attempts to make this situation more intuitively palatable, from reconceiving the moment of measurement as a splitting of the world into multiple copies [8], to introducing explicitly random sub-layers of novel physical process [9, 10]. But, in one way or another, all these attempts have in common an axiomatic treatment of discontinuous measurement, random state selection and Born-rule probability, so it's not clear what's really gained conceptually.

The cleanest outcome would be to eliminate the apparent reliance on explicit measurement axioms altogether, either by refuting them experimentally or deriving them in some form from unitary dynamics. The challenge of doing so is known as the quantum measurement problem. Recent developments underscore the problem. History [11, 12] calls into question the original logic behind the conventional measurement axioms. The concept of quantum non-demolition measurement [13] – important for engineering quantum systems – creates a distinction between projective and non-projective measurements that wasn't foreseen by the framers and confounds the conventional axiomatic formulation. Proton-decay experiments [14] search for events with probability many orders of magnitude smaller than in any direct test of the Born rule to date. And quantum computers [15] promise to perform measurements in quantities so voluminous that even small departures from the Born rule could conceivably leave subtle statistical signatures that could bias elaborate calculations.

If there are no universal measurement axioms, then the outcomes of quantum experiments must be understood case-by-case, although there can be recurring themes. In Reference [16], I examined a specific but simpler version of the quantum measurement problem, known as the Mott problem. The Mott problem asks: Is there a microscopic physical mechanism – based (explicitly or implicitly) only on Schroedinger's equation – that explains why a single alpha particle emitted in a single spherically symmetric s-wave nuclear decay produces a manifestly non-spherically-symmetric single track in a cloud chamber? I formulated such a physical mechanism and derived a Born rule for the starting locations of alpha particle tracks from radioactive decay in a cloud chamber. In References [17, 18], I supported my findings with publicly available, if circumstantial, video data. The Born rule followed from the existence of sites – almost-critical condensed clusters of supercooled vapor molecules – at which the cross section for ionization by a passing charged particle is near singular at ionization threshold. Cross section governs the rate at which alpha-particle wavefunction square-norm is siphoned from the original emitted s-wave into a wavefunction corresponding to ionization *at that particular* site. A nearly singular cross section means that the rate at which square-norm flows into the ionization channel *at that particular* site is much larger than at other sites. So ionization at that particular site consumes the bulk of the original emitted s-wave wavefunction at the expense of all other sites.

In Reference [16], the randomness of the detection location (i.e. the origin of the cloud chamber track) was not intrinsic to quantum mechanics, but reflected the statistical mechanics of vapor condensation. The near-singularity arose from a Penning-like process involving molecular polarization in the cluster. In what follows, for reasons that will become clear, I refer to this threshold near-singularity enabled by induced collective polarization as "leveraged ionization."

The cloud chamber is of great pedagogical and heuristic significance because one doesn't have to imagine condensed vapor clusters: when they're big enough, one can see them with the naked eye in the aggregate as vapor trails, and under low-power magnification as individuated little spherical droplets. But we are left wondering how to explain the probabilistic nature of other techniques for detecting radioactive decay products – and beyond that, more general types of quantum measurement – that don't involve condensation of supercooled vapor. In the present paper, I examine a specific non-condensation detector, the Geiger counter [19]. In this connection I make four principal innovations: First, I argue that a Mott-problem-like question can and should also be asked about Geiger counters. Second, I formulate a corresponding leveraged-ionization mechanism for Geiger counters based on solid-state defects in the Geiger counter window. This parallels the mechanism introduced in Reference [16] for cloud chambers and accomplishes much the same thing. Third, I define a very simple experiment to see whether it's really true that the same ionization physics that takes place in cloud chamber subcritical vapor clusters also occurs at defects in the thin mica window of a Geiger counter. Fourth, although I do not have a microscopic theory of window defects, I report on a proof-of-concept execution of the experiment, and present results that support the leveraged-ionization picture.

These results are significant in their own right, but perhaps even more so because similar defect-driven ionization physics could take place in other solid-state media that figure prominently in instances of the full quantum measurement problem. For this reason, in two corollary innovations, I show how such physics in solid-state detectors and analog-to-digital converters could account for canonical quantum measurement outcomes in Stern-Gerlach experiments and in a particular realization of superconducting-qubits.

The remainder of this paper is organized as follows: Section 2 establishes a conceptual baseline by reviewing the cloud chamber picture from References [16-18]. Section 3 contains our discussion of Geiger counters. Section 4 shows how to apply the concept of leveraged ionization to the Stern-Gerlach experiment, and Section 5 does the same thing for a particular realization of superconducting qubits. Section 6 summarizes our work, including a roadmap for future investigation.

## 2. Cloud chamber

This section (which includes paraphrased content from Reference [18]) summarizes the main points of the theory of cloud chambers developed in References [16-18]. It is necessary context for the remainder of this paper, but is in no way intended to stand alone as a derivation – rigorous or qualitative – of the basic results, or to add anything new to our understanding of cloud chambers. The reader is referred to References [16-18] for more fundamental treatments.

A diffusion cloud chamber is an enclosure containing air commonly supersaturated with an ethyl alcohol vapor. A passing charged particle ionizes air molecules, and the ions nucleate visible vapor droplets. This picture is easy to understand when the charged particle wavefunction is strongly collimated (so the particle can be treated as a point at one location moving in one direction). But the actual wavefunction of an alpha particle near the source of an s-wave radioactive decay is spherically symmetric and not collimated in any meaningful sense.

To *initiate* the well-defined track seen in experiments, the diffuse alpha wavefunction interacts with a vapor cluster (generated randomly by thermal fluctuations) that is just barely subcritical. A barely sub-critical cluster turns out to have a very large cross section for ionization by the alpha, so that even a very weak alpha wavefunction can provoke the subcritical droplet to grow quickly in a supercritical fashion and become visible (at the expense of ionization at other sites, as explained in Section 1), and also provoke the alpha wavefunction to collimate into a narrow beam originating at the point of ionization. The final-state alpha wavefunction is collimated because in the course of ionization it's effectively "reradiated" by a source (the ionized atom) of atomic size, much larger than the alpha wavelength. Once the alpha wavefunction is collimated, all subsequent ionizations must be along the (roughly) straight-line path defined by the beam of collimation. The ionization cross section of a barely sub-critical cluster is very large because, in a vapor cluster, single-molecule ionization can proceed with very small energy loss: the ion can induce (negative) potential energy due to cluster polarization that can nearly balance the (positive) energy needed for the alpha to excite the ejected electron (a form of Penning ionization [20]). I.e., ionization is levered by induced collective polarization. This near-degeneracy drives the cross section of this quantum Coulomb interaction to near singularity. In Reference [16] it is argued that the singularity that approximates the cross section takes the specific form

$$\sigma \sim \frac{A}{R_c - R}, \tag{2.1}$$

for $R$ close to $R_c$, where $R$ is cluster radius and $R_c$ is a critical value, and $A$ is a constant characteristic of the ionization process (the ingredients that go into $A$ are discussed in detail in Ref. [18]). The condition for a track to successfully form at a specific cluster is

$$A s \tau |\psi|^2 > R_c - R, \tag{2.2}$$

where $s$ is alpha particle speed (momentum divided by mass), $\tau$ is cluster evaporation lifetime and $\psi$ is alpha wavefunction. [We can separate $\tau$ and $|\psi|^2$ in this way because for a slow decay, the decay product is in a Gamow state [21] and $|\psi|^2$ in the detector varies slowly with time.] If $\rho dR$ is the number of clusters per unit volume and unit time with radius between $R$ and $R+dR$, then the probability of track initiation per unit volume and time is

$$\rho A s \tau |\psi|^2. \tag{2.3}$$

For a point radiation source, this depends on distance $r$ from the source as

$$\frac{C}{r^2} \tag{2.4}$$

for some position-independent coefficient $C$.   (Equation (2.4) only applies inside the cloud chamber, so one need not be concerned that, formally, Equation (2.4) results in a divergence when integrated volumetrically out to infinitely large distance.)

Bear in mind that clusters are not in fact continuous media, but are made of discrete molecular units, so there can actually be a minimum attainable value of $R_c$-$R$; this is one of presumably many ways in which the singularity in Eq. (2.1) can break down.   That can make the Born rule, in the form of Equations (2.3) and (2.4), break down for small enough $\psi$.   This possibility is discussed in greater detail in [17].

It's easy to extend the argumentation in [16] to the case of a moving charged particle wave packet (group velocity = $s$) that crosses the ionization site in time less than $\tau$.   Then the combination $s\tau|\psi|^2$ in Equations (2.2) and (2.3) would be replaced by

$$\int_{-\infty}^{+\infty} |\psi|^2 \, dz, \tag{2.5}$$

where the integration variable $z$ is distance along a path that runs through the ionization site along the particle's group velocity.   Moreover, Equation (2.3) would then refer to a probability per unit volume but not per unit time, and $\rho dR$ would need to refer to the number of clusters with radius between $R$ and $R+dR$, per unit volume, but not per unit time.   This will be relevant in Sections 4 and 5.

[The reader may find it tangentially interesting to note that decoherence, which figures in the general literature on quantum-mechanical density matrices [22], plays no explicit role in References [16-18].   Presumably, decoherence appears implicitly by proxy via the thermal fluctuations that produce random ensembles of sub-critical vapor clusters.]

## 3. Geiger counter

We can now turn to the four main innovations enumerated in Section 1, which we can encapsulate under the following umbrella statement: The Mott problem is not a narrow question confined to one particular detection system – cloud chambers – but is actually broader, pertaining at least also to Geiger counters.

The heart of a Geiger counter is a metallic (Geiger-Muller – GM) tube containing a noble gas (plus inert buffer), with an electrified wire running down the center, sealed by a thin mica window at one end (the entrance) and metal at the other.   A charged particle passing through the tube, entering through the window, ionizes a trail of gas atoms, and the resulting ions and electrons trigger a current avalanche whose arrival at the wire and the tube walls produces an observable voltage pulse.   There are no thermal-fluctuation-driven clusters in the far-from-supersaturated gas.   So if the charged particle to be detected arrives at the window as a spherically symmetric wavefunction emitted by radioactive decay, there is nothing in the gas inside (or in the air outside) the tube to trigger wavefunction collimation and initiate a track.   But alpha emitters since time immemorial have been placed close to Geiger counter windows, producing many detection counts, so wavefunction collimation must happen *somewhere else*. [Such a statement has never been made before and constitutes our first innovation.]   If it's not in air or in the tube gas, the next places to look are inside the source itself, *or inside the solid window* at hypothetical sites (presumed crystal defects with possibly fluctuating parameters) where, as in cloud chambers, ions can induce just the right amount of (negative) potential energy – by polarizing the surrounding medium – to cancel the (positive) energy of ionization. ($R_c$ in this case becomes a critical defect size, and I ignore any complication in the distribution of defect shapes that undercuts using $R$ and $R_c$ as meaningful parameters.) [Hypothesizing such sites is our second innovation.]

I do not have a microscopic theory of such defects. Instead, I propose probing them indirectly through an experiment in which a point-like alpha source is placed near a Geiger counter window, and count rate is measured as a function of separation between source and window, i.e. between decay source and the hypothetical locus of wave-function collimation from spherical to track-like. One can easily derive (see below) the form that this function *should* take if the underlying process is governed by Equation (2.4) in the window. If this derived form fits poorly, that rules out the leveraged ionization picture. [The identification of this experiment concept is the third innovation.]

In May 2022, I conducted a test of this experimental concept (Figure 1). I mounted a needle source [23] of the alpha emitter $^{210}$Po on a manual analog linear stage calibrated in fractions of an inch. I dialed the stage until the "hot" end of the needle just touched the center of the 16-mm-diameter mica window of a commercial Geiger-Muller tube [24] (I first exposed the window by removing a protective wire mesh). I then dialed the stage back by multiples of 1/40 in = 0.64 mm, and, at each multiple, I watched the analog dial of an activity meter [25] and recorded the lowest value of counts per minute (CPM) attained during a 30 sec interval (this minimum was much easier to judge by eye than the mean value). The background was 50 CPM (compare with 40 CPM per the detector spec sheet [24]). For each stage setting, I subtracted this background from the recorded counts per minute with the source in place, and then corrected the resulting values ($N$) by adding a simple scaled square root to back out what the *average* count rates should be. I used the specific scaled square root $(15N)^{1/2}$ because the meter is hardwired to integrate over a moving time interval of 4 sec and there are fifteen such integration periods in a minute. This is how I arrived at the values on the horizontal axes in Figures 3 and 4 below.

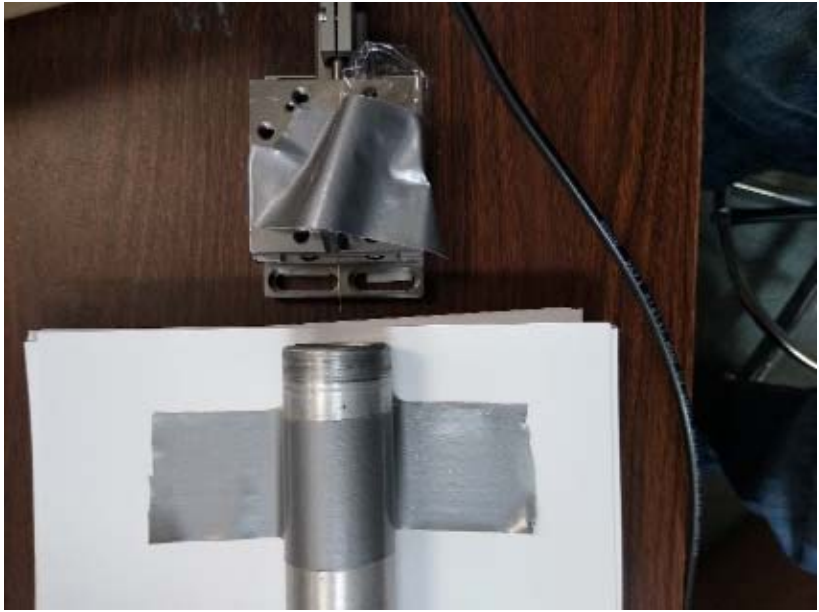

**Figure 1.** Test of experimental concept. The Geiger-Muller tube is the cylinder in the bottom half of the photograph. One can make out the very thin source needle issuing from the plug at the lower edge of the movable stage in the top half of the photograph.

To analyze the results, I compare the measurements with two models. One model is naively geometric; it assumes that spherical wavefunctions convert to collimated wavefunctions in the source medium, so that detector CPM is the total source activity into 4π steradians *multiplied by* the fraction of solid angle subtended by the mica window, but cut off as dictated by alpha stopping power in air and mica. The other model assumes that track collimation takes place inside the window according to the Born rule stated in Equation (2.4), again cut off as dictated by alpha stopping power in air and mica.

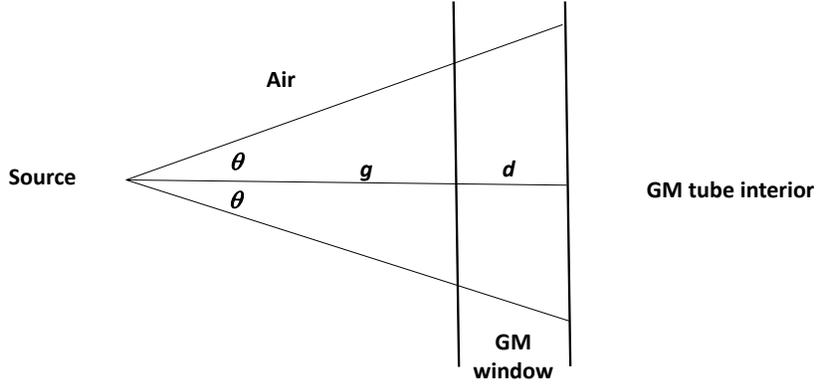

**Figure 2.** Illustration of model correction for alpha stopping.

The stopping correction for an ideal point source is diagrammed in Figure 2. (In actuality, the needle source is about 4 mm long [17], I correct for that in Figure 4.) The parameter $d$ is the thickness of the window *scaled to the distance in air that gives equivalent alpha stopping* (the actual unscaled thickness is small enough to be irrelevant for now). As a practical matter, I do not actually know the precise value of $d$ a priori and have to fit it to the data, as appropriate (see below). The parameter $\theta$ is the smaller of the angle of a ray that just intersects the window's edge, and the angle of a ray whose total length in the diagram is equivalent to the alpha-particle extinction length $L$ in air (roughly 4 cm for $^{210}$Po alphas [26]). In either case, real or virtual alphas corresponding to rays with angles greater than $\theta$ cannot produce ion trails in the GM tube. In formulas, we have

$$\theta = min\left\{tan^{-1}\left(\frac{D}{2g}\right), cos^{-1}\left(\frac{g+d}{L}\right)\right\}, \tag{3.1}$$

where $D$ is window diameter. The count rate in the geometric model is then the total source activity $F$ multiplied by the fraction of solid angle subtended by a cone of angle $\theta$, i.e.

$$\left(\frac{F}{2}\right)(1-cos\theta). \tag{3.2}$$

For the radioactive source that I used, $F$ = (0.01±20%) μCurie = 22,200±20% CPM (not accounting for possible degradation since several weeks elapsed between when I acquired the source and when I used it; that could only worsen the fit between experimental data and the geometric model – see below).

The count rate in the model based on the cloud-chamber Born rule is derived from the integral of Equation (2.4) over the interior of the window, cut off by angle $\theta$. The integral, using polar coordinates in the plane of the window, is

$$\int dz \int_0^{(g+z)\tan\theta} \left(\frac{C}{h^2 + (g+z)^2}\right) 2\pi h\, dh = -2\pi CZ \ln(\cos\theta) \equiv -C' \ln(\cos\theta), \quad (3.3)$$

where $z$ is non-scaled distance into the window, and $Z$ is non-scaled total window thickness.

The model comparisons for a point source idealization are shown in Figure 3.

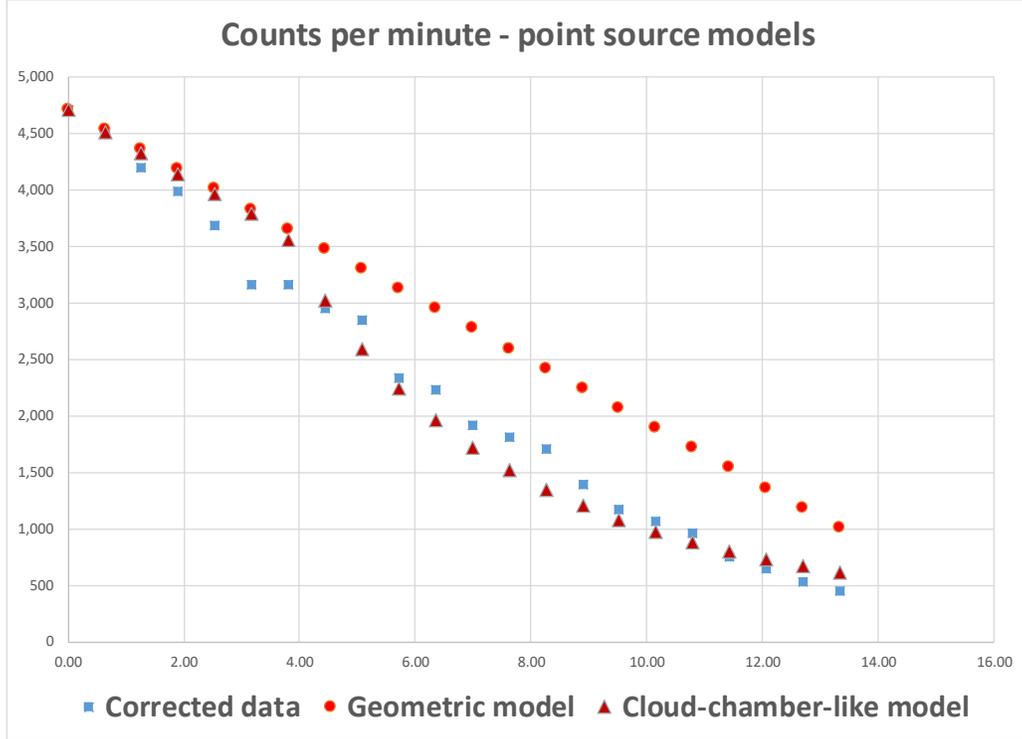

**Figure 3.** Data and ideal point-source models for Geiger counter experiment. Horizontal axis is source-window separation in mm. Vertical axis is counts per minute. Red circles correspond to geometrical model, purple triangles to cloud-chamber-like model, and blue squares to background-subtracted data corrected to provide mean rather than minimum value over 30 sec observation.

The parameter $d$ in the geometric model is fixed by total source activity and normalizing to the corrected experimental count rate 4708 CPM at zero separation,

$$d = (4\ cm)\left(1 - \frac{9416}{F}\right). \quad (3.4)$$

This model comes as close as it can to the rest of the data points when $F$ is as large as possible, i.e. 1.2 μCurie, whence $d=23$mm. In the alternative cloud-chamber-inspired model, the parameter $d$ is set by hand and the coefficient $C'$ is then fixed by $d$ and the experimental count rate at zero separation. The best fit achievable in this case has $d=14$mm. The cloud-chamber-inspired model seems to provide a much better fit to the data than the geometric model.

The comparisons for extended-source versions of the models are shown in Figure 4. To get each model point in Figure 4, I averaged each point-source model value over five successive values of source-window separation, and then adjusted parameters for best fit. Best fit for the geometric model again has total activity of 1.2 μCurie, but this time with $d=22$mm. Best fit for the model based on Equation (2.4) has $d=16$mm. Again the cloud-chamber-like model seems to provide a much better fit to the data than the geometric model, and the extended-source model seems to provide a better fit than the point source model. Thus we may have the beginning of an empirical existence proof for leveraged ionization playing a basic role in Geiger counter behavior.

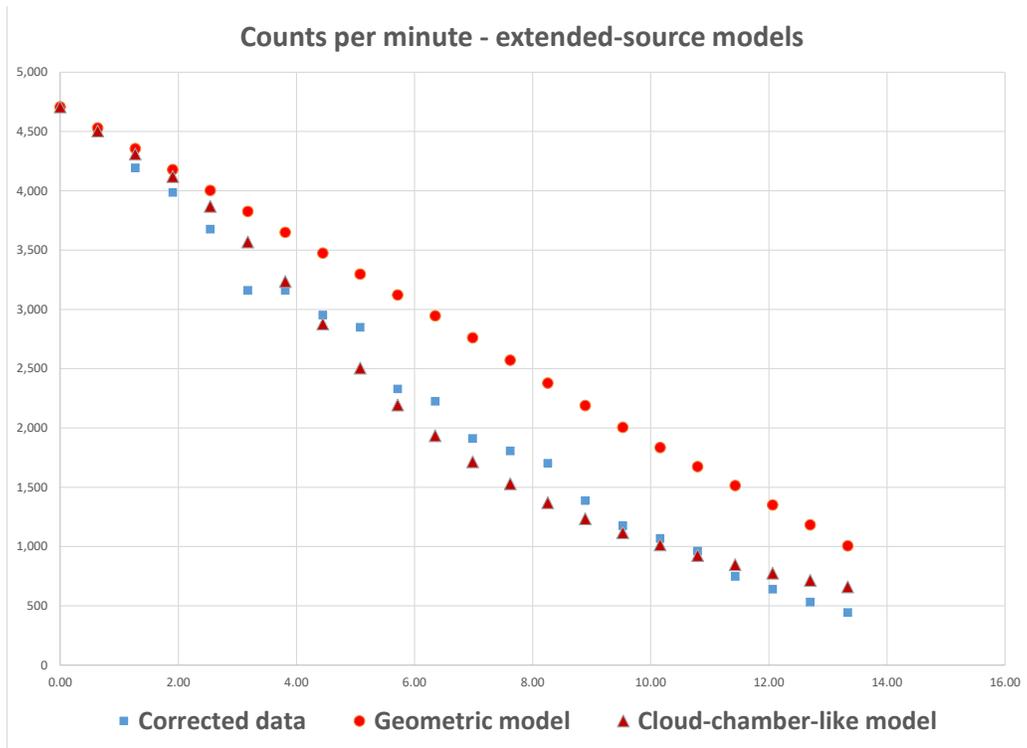

**Figure 4.** Data and ideal extended-source models for Geiger counter experiment. Red circles correspond to geometrical model, purple triangles to cloud-chamber-like model, and blue squares to background-subtracted data corrected to provide mean rather than minimum value over 30 sec observation.

The fit value of $d$ is actually an independent check on these results. According to Fig. 3 in [27], for an alpha particle from $^{210}$Po decay (5.4 MeV, 1.6x10$^9$ cm/s), 1 cm of air has the same stopping power as 1.44 mg/cm$^2$ of mica. In this experiment's Geiger counter [24], the window is rated at 2.0±0.3 mg/cm$^2$. That means $d$=(2.0±0.3)/1.44 cm = 14±2 mm, in line with the cloud-chamber-model results.

The results represented in Figures 3 and 4 constitute our fourth innovation.

A more detailed model might take into account the GM tube's metallic wall or central wire, but nothing in these results appears to cry out for including such effects. This is perhaps to be expected, since wall and wire are much farther than window from the source.

These results seem at odds with student-level demonstrations of an inverse-square law for detections from a (typically only beta or gamma) radioactive source [28], because the inverse-square law is essentially the geometrical model disfavored above. I can only suppose that the source media used in such demonstrations are thick enough to contain plenty of their own sites for spherical-to-collimated wavefunction conversion. By contrast, the radioactive film at the tip of the needle source here must not be thick enough, leaving the Geiger counter window to pick up the slack.

## 4. Stern-Gerlach experiment

If the Mott problem extended to a Geiger counter really is an existence proof for leveraged ionization playing a central role in a simpler version of the quantum measurement problem, then maybe it plays a similar role in instances of the *full* quantum measurement problem with charged particles and solid-state instrumentation. With this hunch in mind, in this section we show how a leveraged-ionization mechanism could account for canonical quantum measurement outcomes in the particular case of the Stern-Gerlach experiment, and in the next section we attempt the same thing for a particular realization of superconducting qubits.

In the original Stern-Gerlach experiment [29, 30], a collimated beam of spin-1/2 neutral silver atoms passed through a region with a spatially varying magnetic field. The beam was then detected as two distinct smears deposited on a glass plate. Later experiments (e.g., Rabi oscillations in Fig. 2A of [31]) refined the experiment to use electronic detectors and verify the Born rule. In what follows we restrict ourselves to a variant in which the spin-1/2 beam particles are actually charged, so that we can apply the results from the preceding section without adding physical complexity beyond the scope of this paper. [The Stern-Gerlach experiment is actually much more difficult practically with a charged-particle beam, and as late as 2019 had not been carried out definitively [32].]

If leveraged ionization occurs in the detector, then the probability per unit area of a charged beam particle hitting the detector of a Stern-Gerlach apparatus at a particular coordinate "**x**" perpendicular to the beam direction is

$$\rho A \int_{-\infty}^{+\infty} |\psi(\pmb{x},z)|^2 \, dz \equiv \rho A \int_{-\infty}^{+\infty} |\psi_{up}(\pmb{x},z)|^2 + \rho A \int_{-\infty}^{+\infty} |\psi_{down}(\pmb{x},z)|^2, \qquad (4.1)$$

where $\psi$ is the incoming two-component wavefunction, $\psi_{up}$ and $\psi_{down}$ are the spatial wavefunctions of the two incoming spin components, and we assume the ionization process is insensitive to spin. (We have also assumed that the nearly singular cross-section as parametrized in Equation (2.1) is large enough to encompass the combined transverse spreads of both up and down components of $\psi$.) If the incoming beam particle is prepared as $a$|spin-up>+$b$|spin-down>, where |spin-up> and |spin-down> are equivalently normalized, then, trivially, we can write the probability in Equation (4.1) as

$$|a|^2 \Phi_{up}(\pmb{x}) + |b|^2 \Phi_{down}(\pmb{x}), \qquad (4.2)$$

for some equivalently normalized functions $\Phi_{up}$ and $\Phi_{down}$. If the $\Phi_{up}$ and $\Phi_{down}$ spots don't overlap, then the probabilities of landing in those two spots are

$$|a|^2 \int \Phi_{up}(\pmb{x}) \, d^2x \text{ and } |b|^2 \int \Phi_{down}(\pmb{x}) \, d^2x$$
$$= |b|^2 \int \Phi_{up}(\pmb{x}) \, d^2x, \qquad (4.3)$$

where the last equality comes from equivalence of normalization. This is the Born rule for spin.

## 5. Superconducting qubit

We are interested in the superconducting qubit here because it represents a step up in sophistication, adding an instructive twist to how leveraged ionization is applied in Section 4. In the Stern-Gerlach experiment, it seemed intuitively obvious that if leveraged ionization that drives canonical quantum measurement outcomes takes place anywhere, it must be at the beam detector. In the qubit case, the locus of decisive leveraged ionization is less obvious intuitively, and so this application requires the additional step of identifying exactly where that locus is. This is likely to be true for the majority of realistically complex quantum measurement scenarios.

A superconducting qubit is an artificial atom made from Josephson junctions coupled to an RF resonating cavity [33]. It is typically configured so that the lowest two excited states |g> and |e> are close in energy and can be treated together as a self-contained two-level system.

In a particular implementation[1] [33], the state of this two-level system is measured ("read out") by sending a microwave pure-tone at the cavity via a transmission line, and recording the reflected signal. If the frequency of the pure tone is chosen appropriately, the signal reflected from an only-|g> qubit has a phase shift that is detectably different from the phase shift due to reflection from an only-|e> qubit. The reflected signal goes through several stages of amplification and then passes through an A/D converter, after which it is recorded as a digitized voltage time series. The phase shift is extracted from the time series via traditional I/Q processing, and the measured state is inferred directly from the result. As a practical matter, phase shifts not corresponding to only-|g> or only-|e> are not observed, since in practice the outcome appears to accord with the canonical measurement axioms: measurement seems to collapse a linear combination $a|g>+b|e>$ randomly to one or another of the two constituent energy levels, with Born-rule probabilities.

This complex setup involves many parts acting over a nontrivial time period [34]: Josephson junctions, a resonator, the machinery of initial state preparation (strings of Rabi pulses), a transmission line, several levels of amplification, A/D conversion, digitized signal recording, and computational processing. If the canonical measurement outcome reflects some underlying decisive physical process founded on Schroedinger's equation, there is a surfeit of places where that process might actually happen. In this section I hypothesize the simplest option: all that matters to the canonical nature of the outcome are the initial qubit state and the final digitization of the reflected and amplified voltage signal, because the A/D converter is the first place along the signal chain where I can imagine that discrimination between only-|g> and only-|e> can take place. (I recognize that my imagination is no substitute for rigorous proof.)

In one of the simplest designs [35], A/D conversion starts by passing a signal through a multi-level voltage divider to generate multiple scaled copies; then it runs the copies through comparators (basically diodes) that output saturated digital 0 or 1 pulses depending on whether the voltages are larger or smaller than some built-in reference values. To get at the essential physics, I (over)simplify the divider network and comparator bank to be a single slab (Figure 5). In this idealization, the signal is manifested as a voltage difference between the slab's top and bottom faces; and a reservoir of mobile electrons occupies a thin planar sheet midway between top and bottom. I imagine that electron transport in the slab is observed at a time when only-|g> and only-|e> voltages are expected to have known and opposite signs. This is conceivable – at least for large enough phase shifts – because the entirety of the only-|g> and only-|e> signals are in principle known and predictable from the timing of the tone generator, together with the configuration of the qubit/resonator system, the amplifier delays, and the signal path, and correspondingly tunable.

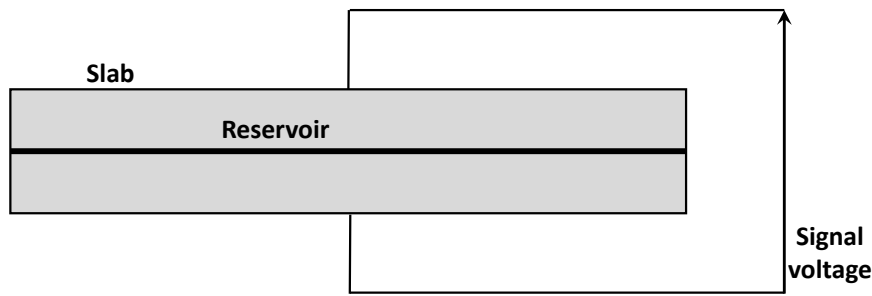

**Figure 5.** Idealized A/D converter.

When a single electron is in the reservoir, it's entangled with the prepared qubit state $a|g>+b|e>$; below the reservoir, it's entangled only with $a|g>$; above, it's entangled only with $b|e>$. This is virtually identical to the Stern-Gerlach case, with |g> and |e> taking the roles of

---

[1] A more comprehensive treatment of qubits more generally is beyond the scope of this paper.

|spin down> and |spin up>. So, if leveraged ionization takes place in the slab, then the Born rule for this two-level system falls out in the same way as discussed in Section 4 for a Stern-Gerlach apparatus.

## 6. Conclusion

I have argued for the novel hypothesis that the behavior of a Geiger counter in the presence of a slow radioactive decay generalizes an earlier account of the Mott problem in a cloud chamber. Although I do not have a microscopic theory of relevant defects in the Geiger counter window, I have proposed an experiment to investigate this generalization, and have presented supporting results from an early proof of concept. I have further argued for the novel corollary that similar physics could account for the *full* quantum measurement problem in at least the charged-beam version of the Stern-Gerlach experiment, and in a particular realization of superconducting qubits. If the ideas in this paper are borne out by further work, it opens the possibility of quantitatively characterizing when canonical measurement behavior breaks down and how that limits quantum computing technology, or even one's ability to measure the lifetime of the proton.

An alternative way to test the track-initiation physics of Geiger counter windows might be to observe alpha tracks in a cloud chamber with a Geiger counter window (*just* the window, separated from the rest of the counter) inserted very close to a small radioactive source. One might expect to see a preponderance of tracks originating at the window itself, rather than at locations elsewhere in the cloud chamber medium.

Similar physics could operate in photographic film, where a photon excites an electron into the conduction band of a silver-halide grain, eventually to be captured by a trap composed of silver atoms [36]. The conduction-band electron wavefunction is spread over the entire grain, but it may need collimation to be trapped effectively, and perhaps that's accomplished by a site of leveraged ionization. Something similar might also hold for electrons in a CCD pixel.

(If leveraged ionization sites somehow existed in liquids, we could calculate limits on the ability of large water-based experiments to detect proton decay [14] at lifetimes of current theoretical interest.)

This picture could possibly also shed some light on a pervasive but unquestioned aspect of our everyday world, namely, that almost all particles we observe seem to follow tracks. Perhaps the wavefunctions of particles we deal with every day become collimated by interacting with leveraged ionization sites in ambient condensed matter. (I first pointed out the lack of a practical theory of particle wavepackets in the real world in Reference [37].)

It is obvious that further work is needed. This includes:
1. A carefully controlled cloud chamber experiment to measure how the origins of tracks from radioactive decay are distributed (making rigorous the data in [17]).
2. A carefully controlled cloud chamber experiment in which a Geiger counter window is placed very close to a point source of alpha decay.
3. A carefully controlled repeat of the experiment described in Section 3.
4. A microscopic theory of leveraged ionization in solids and liquids, with particular attention to Geiger-counter windows, real analog-to-digital converters (not just the toy model of Section 5), photographic grains, CCD pixels, and water.
5. A fully rigorous treatment of the singular form in Eq. (2.1).
6. A careful analysis of the lower limits to $|R_c-R|$ at a leveraged ionization site. This can drive limits to the Born rule.

**Acknowledgments**

I am grateful to Thomas Gauron, Edward Hertz and Janice Pacenka (Harvard-Smithsonian Center for Astrophysics) for access to equipment and facilities; to Rhonda Harris (Ludlum Measurements Inc.) for answering detailed questions about Geiger counter operation; to Daniel Sims (Spectrum Techniques Inc.) for answering detailed questions about the $^{210}$Po needle source; to Ron Folman and Yair Margalit (Ben-Gurion University), and Horst Schmidt-Bocking (University of Frankfurt) for correspondence about the Stern-Gerlach experiment; and to Will Oliver and Max Hays (MIT) for educating me about superconducting qubits.

**Author Contributions**

The single author of this paper is solely responsible for its content.

**Competing Interests**

The author declares no competing interests

**Funding**

This research received no external funding.

**Data Availability Statement**

Data available upon request.

## References

1. Von Neumann, J. *Mathematical Foundations of Quantum Mechanics*. Princeton University Press, Princeton, 1955.
2. Aspden, R., Padgett, M., Spalding, G. Video recording true single-photon double-slit interference. Am. J. Phys. 2016; 84: 671.
3. Tonomura, A. et al. Demonstration of single-electron buildup of an interference pattern. Am. J. Phys. 1989; 57: 117.
4. Zeilinger, A. et al. Single- and double-slit diffraction of neutrons. Rev. Mod. Phys. 1988; 60: 1067.
5. Hackermuller, L., et al. Decoherence of matter waves by thermal emission of radiation. Nature 2004; 427: 711.
6. Gerlich, S. et al. Quantum interference of large organic molecules. Nature Communications 2011; 2: 263.
7. Bialecki, T., Ryboticki, T., Tworzydlo, J., Bednorz, A. Born rule as a test of the accuracy of a public quantum computer. arXiv:2112.07567v3 [quant-ph] 2021.
8. Wallace, D. Everett and structure. Studies in History and Philosophy of Modern Physics 2003; 34: 87–105.
9. Holland, P. R. The Quantum Theory of Motion: An Account of the de Broglie-Bohm Causal Interpretation of Quantum Mechanics. Cambridge University Press, 1995.
10. Bassi, A. and Ghirardi, G. Dynamical reduction models. Physics Reports 2003; 379: 257–426.
11. Sen, R. N. Homer nodded once more. Von Neumann's misreading of the Compton-Simon experiment and its fallout. arXiv:2302.14610 [quant-ph] 2023.
12. Freericks, J. K. How to measure the momentum of single quanta. arXiv:2302.12303 [quant-ph] 2023.
13. Brun, T. A. A simple model of quantum trajectories. Am. J. Phys. 2002; 70: 719.
14. Wikipedia. Proton decay. https://en.wikipedia.org/wiki/Proton_decay
15. Krantz, P. et al. A quantum engineer's guide to superconducting qubits. Appl. Phys. Rev. 2019; 6: 021318.
16. Schonfeld, J. F. The first droplet in a cloud chamber track. Found. Phys. 2021; 51: 47.
17. Schonfeld, J. F. Measured distribution of cloud chamber tracks from radioactive decay: a new empirical approach to investigating the quantum measurement problem. Open Physics 2022; 20: 40.
18. Schonfeld, J. F. Order-of-magnitude test of a theory of the Mott problem. arXiv:2209.05344 [physics.gen-ph] 2022.
19. Wikipedia. Geiger-Muller tube. https://en.wikipedia.org/wiki/Geiger%E2%80%93M%C3%BCller_tube.

20. Efimov, D. K., Miculis, K., Bezuglov, N. N., Ekers, A. Strong enhancement of Penning ionization for asymmetric atom pairs in cold Rydberg gases: the Tom and Jerry effect. J. Phys. B: At. Mol. Opt. Phys. 2016; 49: 125302.
21. Garcia-Calderon, G. and Peirels, R. Resonant states and their uses. Nucl. Phys. 1976; A265: 443.
22. Schlosshauer, M. A. *Decoherence and the Quantum-to-Classical Transition*. Springer, Berlin, 2007
23. https://www.spectrumtechniques.com/product/pb-210-needle-source/
24. https://ludlums.com/products/all-products/product/model-44-7
25. https://ludlums.com/products/all-products/product/model-3
26. Sims. D (Spectrum Techniques Inc.), private communication, 2023.
27. Bennet, W. E. The stopping power of mica for $\alpha$-particles. Proc. Roy. Soc. A 1936; 155: 419.
28. https://www2.lbl.gov/abc/experiments/Experiment1.html
29. Gerlach, W. and Stern, O. Der experimentelle Nachweis der Richtungsquantelung im Magnetfeld. Z. Phys. 1922; 8: 110.
30. Bauer, M. The Stern-Gerlach Experiment Translation of: "Der experimentelle Nachweis der Richtungsquantelung im Magnetfeld". arXiv:2302.11343 [physics.hist-ph] 2023.
31. Zhou, Z. et al. An experimental test of the geodesic rule proposition for the noncyclic geometric phase, Sci. Adv. 2020; 6: eaay8345.
32. Henkel, C. et al. Stern-Gerlach splitting of low-energy ion beams. New J. Phys. 2019; 21: 083022.
33. Blais, A. et al. Cavity quantum electrodynamics for superconducting electrical circuits: an architecture for quantum computation. Phys. Rev. 2004; A69: 062320.
34. Peronnin, T., Markovic, D., Ficheux, Q., Huard, B. Sequential dispersive measurement of a superconducting qubit. Phys. Rev. Lett. 2020; 124: 180502.
35. Wikipedia. Flash ADC. https://en.wikipedia.org/wiki/Flash_ADC
36. Belloni, J. The role of silver clusters in photography. C. R. Physique 2002; 3: 381.
37. Schonfeld, J. F. Analysis of double-slit interference experiment at the atomic level. Studies in History and Philosophy of Modern Physics 2019; 67: 20.